\documentstyle[prl,aps,preprint,tighten]{revtex}

\begin{document}
\draft


\preprint{\vbox{\it Submitted to Phys.\ Rev.\ D --- Brief Reports \hfill\rm
DOE/ER/40762--038\\
                \null\hfill UMPP \#94--142}}
\title{Relations among zero momentum correlators
\linebreak for heavy-light systems in QCD}
\author{Xuemin Jin and Thomas~D. Cohen}
\address{Department of Physics and Center for Theoretical Physics\\
University of Maryland, College Park, Maryland 20742}
\date{\today}
\maketitle
\begin{abstract}
Relations connecting various zero momentum correlators of interpolating
fields for pseudoscalar and scalar channel, containing one heavy and
one light quark field, are derived from the Euclidean space formulation
of the QCD functional integral. These relations may serve as
constraints on the phenomenological models or approaches motivated from
QCD, and suggest a method to extract the chiral quark condensates.  It
is also found that the correlator for pseudoscalar channel differs from that
for scalar channel even in the large heavy quark mass limit.
\end{abstract}

Heavy-light systems made of one heavy and one light quark have been
attracting attentions, motivated by the observation of heavy quark
symmetry in the infinite heavy quark mass limit\cite{isgur1}.  The
basis of this symmetry is a separation  between the scale of the heavy
quark mass and all other dynamical scales in the problem.

The properties of heavy-light systems have been investigated using
various phenomenological approaches such as QCD sum-rule
method\cite{shifman1,shuryak1,radyushkin1,neubert1} and numerical
simulations on a lattice\cite{gavela1,bernard1,allton1,alexandrou1,maiani1},
based on the analytic properties of two-point correlators of
interpolating fields, carrying the quantum numbers of the system under
study, and on asymptotic freedom. In these approaches, hadronic
spectral properties ({\it e.g.,} masses, coupling constants, etc.) are
related via dispersion relations to the correlators evaluated in terms
of quark and gluon degrees of freedom.  In practical applications, one
needs to parametrize the spectral functions and to evaluate the
correlators approximately. The success of such approaches depends on
correct understanding of the qualitative features of the spectral
functions and accurate evaluation of the correlators from QCD.

In this paper, we show, within the Euclidean formulation of the QCD
functional integral, that the two-point zero momentum correlators of
interpolating fields for both pseudoscalar and scalar channel
containing one heavy and one light flavor can be expressed as
combinations of quark condensates. We then derive relations which
connect various correlators.  These relations can be regarded as
constraints on phenomenological descriptions of the correlators.
We find that the correlators for pseudoscalar and scalar channel
are different even in the large heavy quark mass limit, suggesting
differences between positive- and negative-parity states.

To evaluate various two-point correlators from QCD, it is necessary to
know the properties of the QCD vacuum. This is, however, difficult due
to the nonperturbative features of QCD at large distances, which lead
to complicated properties of the QCD vacuum such as the formation of
the gluon condensate $\langle 0|(\alpha_s/\pi)G^2|0\rangle$ and the
quark condensate $\langle 0|\overline{q}q|0\rangle$. The latter is of
particular importance as it signals spontaneous breaking of the chiral
symmetry  which the QCD Lagrangian enjoys in the massless limit.

In the pioneering work of Banks and Casher\cite{bank1}, it was
noted that the fermion condensate is closely related to the mean level
density of the fermion eigenvalues $\langle 0|\overline{q}q
|0\rangle\simeq -\pi\rho(\lambda=0)$, where $\rho(\lambda) d\lambda$
denotes the mean number of eigenvalues of the Euclidean Dirac operator,
$\rlap{\,/}{D}$, contained in the interval $d\lambda$, per unit volume.
Thus the formation of a quark condensate is connected with the
occurrence of the unusual behavior of the low eigenvalues of the Dirac
operator, which is dramatically different from that for the free Dirac
operator.

The Bank-Casher formula $\langle 0|\overline{q}q |0\rangle\simeq
-\pi\rho(0)$  can be easily derived in the Euclidean formulation of the
QCD functional integral\cite{leutwyler1,smilga1,smilga2}. In a {\it given}
background gluon field
configuration, the fermion Green's function can be written as\cite{leutwyler1}
\begin{equation}
{\cal G}^f_{\rm A}(x,y)=\langle q_f(x)\overline{q}_f(y)\rangle_{\rm A}=\sum_{n}
{\psi_n(x)\psi^\dagger_n(y)\over m_f-i\lambda_n}\; ,
\label{prop-gen}
\end{equation}
where $f$ is flavor index (a sum over $f$ is not implied), and
$\psi_n(x)$ and $\lambda_n$ are normalized eigenfunctions and
eigenvalues of the Euclidean Dirac operator:
\begin{equation}
\rlap{\,/}{D}\psi_n(x)=\lambda_n\psi_n(x)\; .
\end{equation}
Note that except for zero modes ($\lambda_n=0$), the eigenfunctions
occur in pairs, \{$\psi_n$, $\gamma_5\psi_n$\}, with opposite
eigenvalues, \{$\lambda_n$, $-\lambda_n$\}\cite{leutwyler1,smilga2}.
Consequently, one obtains
\begin{equation}
{1\over V}\int_{\rm V} d^4 x{\rm Tr}\left\{{\cal G}^f_{\rm
A}(x,x)\right\}={1\over V}\sum_{\lambda_n>0}{2m_f\over m_f^2+\lambda_n^2}\; ,
\label{midd}
\end{equation}
where $V$ is the volume of the four-dimensional box introduced to
provide the infrared regularization of the theory and make the spectrum
discrete; at the end, one should take the limit $V\rightarrow\infty$.
The zero-mode contributions, vanishing in the infinite volume limit,
have been droped in Eq.~(\ref{midd}) (see Ref.~\cite{leutwyler1} for details).
Averaging
over all gluon field configurations, one gets for large volumes
\begin{equation}
\langle 0|\overline{q}_fq_f|0\rangle=-2m_f\int_0^\infty d\lambda
{\rho(\lambda)\over m_f^2+\lambda^2}\; ,
\label{bank-cash}
\end{equation}
where $\rho(\lambda)$ is the mean level density. Here the average means
the functional integral over gluon configuration with weighting
function, $e^{-S_{\rm YM}}$ mutiplyed by the fermion determinant
functional, where $S_{\rm YM}$ is the Euclidean action for pure
Yang-Mills theory. Note that the integral in Eq.~(\ref{bank-cash}) is
divergent due to the perturbative ultraviolet contributions.  One can,
for example, regularize this divergence by subtracting the perturbative
contributions from $\rho(\lambda)$ or simulate the functional integral
in a lattice regularized version. In the discussions to follow, we
always assume that the integrals over the mean eigenvalues are properly
regularized. Our results are independent of how the associated
divergences are regularized. Assuming that the quark mass is much less
than the characteristic hadronic scale on which $\rho(\lambda)$ changes
substantially, one immediately reproduces the Bank-Casher formula.

To begin, we consider the correlators of pseudoscalar interpolating
fields containing one {\it heavy} and one {\it light} quark field as
given by
\begin{equation}
\Pi^{\rm PS}_{hl}(q)\equiv i\int d^4x e^{iq\cdot x}\langle 0|{\rm T}j^{\rm
PS}_{hl}(x)(j^{\rm PS}_{hl})^\dagger(0)|0\rangle\; ,
\label{corr-qQ-ps}
\end{equation}
where the colorless interpolating field $j^{\rm PS}_{hl}$ is defined as
\begin{equation}
j^{\rm PS}_{hl}(x)=(m_l+m_h)\overline{l}(x)i\gamma_5 h(x)\; ,
\label{int-qQ-ps}
\end{equation}
where $l(x)$ denotes light quark field ($u$, $d$, $s$) with mass $m_l$
and $h(x)$ is for heavy quark field ($c$, $b$, $t$) with mass $m_h$.
The interpolating field $j^{\rm PS}_{hl}$ has the same quantum numbers
as a pseudoscalar heavy-light meson ($D$, $B$, etc.), and is equal to
the divergence of the axial-vector current,
$A_{hl}^\mu(x)=\overline{l}(x)\gamma^\mu\gamma_5h(x)$.  Since
$A_{hl}^\mu$ is partially conserved, $\Pi^{\rm PS}_{hl}(q)$ is a
renormalization-group invariant quantity. In addition, Lorentz
covariance, parity and time-reversal imply that $\Pi^{\rm PS}_{hl}$ is
a scalar.

Calculation of $\Pi^{\rm PS}_{hl}(q)$ from QCD is in general difficult.
However, at zero momentum, the correlator $\Pi^{\rm PS}_{hl}(0)$ can be
easily evaluated within the Euclidean formulation of the QCD
functional integral.  For a fixed gluon configuration, only the
connected part contributes, and one gets
\begin{eqnarray}
\Pi^{\rm PS}_{hl}(0)\mid_{\rm A}&=&(m_l+m_h)^2\int_{\rm V}d^4x {\rm
Tr}\left\{{\cal G}^h_{\rm A}(x,0)\gamma_5{\cal G}^l_{\rm
A}(0,x)\gamma_5\right\}\; ,
\nonumber
\\*[7.2pt]
&=&(m_l+m_h)^2{1\over V}\int_{\rm V}d^4x d^4y {\rm Tr}\left\{{\cal G}^h_{\rm
A}(x,y)\gamma_5{\cal G}^l_{\rm A}(y,x)\gamma_5\right\}\; ,
\label{expan-med}
\end{eqnarray}
where ${\cal G}^l_{\rm A}$ and ${\cal G}^h_{\rm A}$ have been
defined in Eq.~(\ref{prop-gen}). A simple alegebraic calculation then yields
\begin{equation}
\Pi^{\rm PS}_{hl}(0)\mid_{\rm A}=(m_l+m_h){1\over
V}\sum_{\lambda_n>0}\left[{2m_l\over m_l^2+\lambda_n^2}+{2m_h\over
m_h^2+\lambda_n^2}\right]\; .
\label{expan-fina}
\end{equation}
Averaging over the gluon background and using Eq.~(\ref{bank-cash}), we obtain
\begin{equation}
\Pi^{\rm PS}_{hl}(0)=-(m_l+m_h)\left[\langle 0|\overline{l}l|0\rangle+\langle
0|\overline{h}h|0\rangle\right]\; .
\label{pcac-qQ}
\end{equation}

Thus, the zero momentum correlator $\Pi^{\rm PS}_{hl}(0)$ is related to
the quark masses and  quark condensates. The analytic properties of
$\Pi^{\rm PS}_{hl}(0)$ allow one to write a dispersion representation,
where the spectral function contains the information about {\it all}
the physical states coupling to the interpolating field $j^{\rm
PS}_{hl}$. If $\Pi^{\rm PS}_{hl}(0)$ were saturated by the lowest
physical resonance, one would expect to obtain a relation from
Eq.~(\ref{pcac-qQ}), similar to the Gell-Mann--Oakes--Renner relation,
which relates the properties of the lowest resonance to the quark
masses and quark condensates.  In nature, however, the heavy quark
masses are much greater than the QCD scale parameter, and the
correlator $\Pi^{\rm PS}_{hl}(0)$ is not saturated by the lowest
resonance.

Nevertheless, some useful
relations can be immediately derived from Eq.~(\ref{pcac-qQ}). For two
different light flavors and a fixed heavy flavor, we have
\begin{equation}
\Pi^{\rm PS}_{hl}(0)-\Pi^{\rm PS}_{hl^\prime}(0)=m_h\left[\langle
0|\overline{l^\prime}l^\prime|0\rangle-\langle
0|\overline{l}l|0\rangle\right]\; ,
\label{pcac-indu-1}
\end{equation}
where $l$ and $l^\prime$ are different light flavors and $h$ can be any heavy
flavor. For two different heavy flavors and a fixed light flavor, we get
\begin{equation}
\Pi^{\rm PS}_{hl}(0)-{m_h\over m_{h^\prime}}\Pi^{\rm PS}_{h^\prime
l}(0)=m_h\left[\langle 0|\overline{h^\prime} h^\prime|0\rangle-\langle
0|\overline{h}h|0\rangle\right]\; ,
\label{pcac-indu-2}
\end{equation}
where $h$ and $h^\prime$ are two different heavy flaovrs and
$l$ can be any light flavor. Finally, for two different heavy flavors and two
different light flavors, we find
\begin{equation}
\Pi^{\rm PS}_{hl}(0)-\Pi^{\rm PS}_{hl^\prime}(0)-{m_h\over
m_{h^\prime}}\Pi^{\rm PS}_{h^\prime l}(0)+
{m_h\over m_{h^\prime}}\Pi^{\rm PS}_{h^\prime l^\prime}(0)=0\; ,
\label{pcac-indu-3}
\end{equation}
where $h$ and $h^\prime$ are different heavy flaovrs  and $l$ and
$l^\prime$ are different light flavors. In
Eqs.~(\ref{pcac-indu-1}--\ref{pcac-indu-3}), we have assumed that
$m_{h,h^\prime}\gg m_{l,l^\prime}$ and used $m_{h,h^\prime}\pm
m_{l,l^\prime}\simeq m_{h,h^\prime}$.

Note that in deriving Eqs.~(\ref{pcac-qQ}--\ref{pcac-indu-3}) we have
not made any assumptions, except for $m_{h,h^\prime}\gg
m_{l,l^\prime}$. So, they are all model independent results.  (If one
retains $m_{h,h^\prime}\pm m_{l,l^\prime}$ explicitly in the
derivation, the resulting relations will be exact). If one adopts
dispersion representations for the correlators, these results will
constrain the spectral functions. Phenomenological parametrizations of
the spectral functions, often used in applications such as the lattice
QCD data or QCD sum-rule calculations, must satisfy these constraints.

One notices from Eq.~(\ref{pcac-indu-1}) that
\begin{eqnarray}
\langle 0|\overline{d}d|0\rangle-\langle 0|\overline{u}u|0\rangle &=&
{1\over m_h}\left[\Pi^{\rm PS}_{hu}(0)-\Pi^{\rm PS}_{hd}(0)\right]\; ,
\\*[7.2pt]
\langle 0|\overline{s}s|0\rangle-\langle 0|\overline{u}u|0\rangle &=&
{1\over m_h}\left[\Pi^{\rm PS}_{hu}(0)-\Pi^{\rm PS}_{hs}(0)\right]\; .
\label{su3-bre}
\end{eqnarray}
Therefore, one may determine the SU(3) breaking in light quark condensates by
evaluating the correlators $\Pi^{\rm PS}_{hu}(0)$, $\Pi^{\rm PS}_{hd}(0)$,
and $\Pi^{\rm PS}_{hs}(0)$.

Following the same pattern as used in deriving Eq.~(\ref{pcac-qQ}), one can
easily obtain
\begin{equation}
\Pi^{\rm PS}_{hh}(0)=-4m_h\langle 0|\overline{h}h|0\rangle\; ,
\label{hhps}
\end{equation}
where $\Pi^{\rm PS}_{hh}$ is the correlator of interpolating field $j^{\rm
PS}_{hh}(x)=2m_h\overline{h}(x)i\gamma_5 h(x)$. Combinig this result with
Eq.~(\ref{pcac-qQ}), we have
\begin{equation}
\langle 0|\overline{u}u|0\rangle={1\over m_h}\left[\Pi^{\rm
PS}_{hh}(0)-4\Pi^{\rm PS}_{hu}(0)\right]\; .
\label{qq}
\end{equation}
So, the calculation of $\Pi^{\rm PS}_{hu}(0)$ and $\Pi^{\rm
PS}_{hh}(0)$ will give rise to a determination of the quark condensate
$\langle 0|\overline{u}u|0\rangle$. In the lattice calculations, one
usually determine the quark condensates by extracting $\rho(\lambda=0)$
directly\cite{hamber1,barbour1,kogut1}. It will be interesting to
extract the quark condensates by calculating various correlators
discussed here on a lattice, and compare the results with those
obtained from the direct calculation of the mean level density.

We now turn to the scalar channel. The corresponding correlator is defined by
\begin{equation}
\Pi^{\rm S}_{hl}(q)\equiv i\int d^4xe^{iq\cdot x}\langle 0|{\rm T} j^{\rm
s}_{hl}(x)(j^{\rm s}_{hl})^\dagger(0)|0\rangle\; ,
\label{corr-s}
\end{equation}
where $j^{\rm s}_{hl}(x)=(m_l+m_h)\overline{l}(x)h(x)$ is the scalar
interpolating field. Again, we are interested in zero momentum correlator
$\Pi^{\rm S}_{hl}(0)$, which, in a fixed background gluon configuration, can be
written as
\begin{equation}
\Pi^{\rm S}_{hl}(0)\mid_{\rm A}=-(m_l+m_h)^2{1\over V}\int d^4x d^4y {\rm
Tr}\left\{
{\cal G}^h_{\rm A}(x,y){\cal G}^l_{\rm A}(y,x)\right\}\; .
\end{equation}
Taking into account the pairing properties of the eigenfunctions and completing
the integral, we get
\begin{eqnarray}
\Pi^{\rm S}_{hl}(0)\mid_{\rm A}&=&(m_l+m_h)^2{1\over
V}\sum_{\lambda_n>0}{2\lambda_n^2-2m_l m_h\over
(m_l^2+\lambda_n^2)(m_h^2+\lambda_n^2)}\; ,
\nonumber
\\*[7.2pt]
&=&{(m_l+m_h)^2\over m_h-m_l}{1\over V}\sum_{\lambda_n>0}\left[{2m_h\over
m_h^2+\lambda_n^2}-{2m_l\over m_l^2+\lambda_n^2}\right]\ .
\end{eqnarray}
Averaging over all gluon configurations and using Eq.~(\ref{bank-cash}), one
obtains
\begin{equation}
\Pi^{\rm S}_{hl}(0)={(m_l+m_h)^2\over m_h-m_l}\left[\langle
0|\overline{l}l|0\rangle-\langle 0|\overline{h}h|0\rangle\right]\; .
\label{s-qQ}
\end{equation}
Similarly, one can derive relations which connect different correlators.
Applying Eq.~(\ref{s-qQ}) to two distinct light flavors, one gets
\begin{equation}
\Pi^{\rm S}_{hl}(0)-\Pi^{\rm S}_{hl^\prime}(0)=m_h\left[\langle
0|\overline{l}l|0\rangle-\langle
0|\overline{l^\prime}l^\prime|0\rangle\right]\; .
\label{s-indu-1}
\end{equation}
For a fixed light flavor and two different heavy flavors, one obtains
\begin{equation}
{m_h\over m_{h^\prime}}\Pi^{\rm S}_{h^\prime l}(0)-
\Pi^{\rm S}_{hl}(0)=m_h\left[\langle 0|\overline{h}h|0\rangle-\langle
0|\overline{h^\prime}h^\prime|0\rangle\right]\; .
\label{s-indu-2}
\end{equation}
With two different light flavors and two different heavy flavors, one then
finds
\begin{equation}
\Pi^{\rm S}_{hl}(0)-\Pi^{\rm S}_{hl^\prime}(0)-{m_h\over m_{h^\prime}}\Pi^{\rm
S}_{h^\prime l}(0)
+
{m_h\over m_{h^\prime}}\Pi^{\rm S}_{h^\prime l^\prime}(0)=0\; .
\label{s-indu-3}
\end{equation}
Here we have neglected light quark masses (i.e., $m_{h,h^\prime}\pm
m_{l,l^\prime}\simeq m_{h,h^\prime}$). These equations are model independent,
and can be regarded as constraints on phenomenological descriptions of the
correlators for scalar channel. One may also use Eqs.~(\ref{s-qQ}) and
(\ref{s-indu-1}) to extract the light quark condensates.

Combining Eq.~(\ref{pcac-qQ}) and Eq.~(\ref{s-qQ}), we find the following
relations
\begin{eqnarray}
\Pi^{\rm PS}_{hl}(0)-\Pi^{\rm S}_{hl}(0)&=&-2m_h\langle
0|\overline{l}l|0\rangle\; ,
\label{ps-s-0}
\\*[7.2pt]
\Pi^{\rm PS}_{hl}(0)+\Pi^{\rm S}_{hl}(0)&=&
-2m_h\langle 0|\overline{h}h|0\rangle\; ,
\label{ps-s-1}
\end{eqnarray}
which connect the correlator for pseudoscalar channel to that for
scalar channel. One notices that these two equations can be used to
determine both light and heavy quark condensates. These relations also
show that even in the limit $m_h\rightarrow\infty$, the pseudoscalar
channel differs from the scalar channel.
%
%
It is worth emphasizing that this result does not contradict the heavy
quark symmetry. In the infinite heavy quark mass limit, heavy quark
symmetry implies that the spin and mass of a heavy quark decouple from
the hadronic dynamics; consequently, there are degeneracy in spin, but
not in parity.

{}From Eqs.~(\ref{pcac-indu-1}), (\ref{pcac-indu-2}), (\ref{s-indu-1}), and
(\ref{s-indu-2}), one may also deduce the following relations
\begin{eqnarray}
& &\Pi^{\rm PS}_{hl}(0)-\Pi^{\rm PS}_{hl^\prime}(0)
+\Pi^{\rm S}_{hl}(0)-\Pi^{\rm S}_{hl^\prime}(0)=0\; ,
\label{ps-s-2}
\\*[7.2pt]
& &
\Pi^{\rm PS}_{hl}(0)-
\Pi^{\rm S}_{hl}(0)-{m_h\over m_{h^\prime}}\Pi^{\rm PS}_{h^\prime l}(0)
+{m_h\over m_{h^\prime}}\Pi^{\rm S}_{h^\prime l}(0)=0\; .
\label{ps-s-3}
\end{eqnarray}

In conclusion, we have derived exact relations which connect various
zero momentum correlators of interpolating fields made of one heavy and
one light quark field for pseudoscalar and scalar channel, within
Euclidean formulation of QCD functional integral. These relations may
serve as constrains on phenomenological models for the correlators such
as those used in QCD sum-rule calculations. We also noted that in
principle one can use these relations to extract quark condensates by
calculating the correlators on a lattice.
One can extend the present techniques to study the vector and axial
vector channel.

\vspace{0.5in}

This work is supported by the U.S. Department of Energy under grant No.
DE-FG02-93ER-40762 and U.S. National Science Fundations under grant No.
PHY-9058487.

\end{document}